\documentclass[twocolumn,showpacs,preprintnumbers]{revtex4}%
\usepackage{amssymb}
\usepackage{amsmath}
\usepackage{graphicx}
\usepackage{dcolumn}
\usepackage{bm}
\usepackage{amsfonts}%
\newtheorem{thm}{Theorem}[section]

\numberwithin{equation}{section}

\begin{document}
\preprint{ }
\title{Conceptual Explanation for the Algebra in the Noncommutative Approach to the
Standard Model}
\author{Ali H. Chamseddine$^{1,3}$and Alain Connes$^{2,3,4}$}
\affiliation{$^{1}$Physics Department, American University of Beirut, Lebanon}
\affiliation{$^{2}$College de France, 3 rue Ulm, F75005, Paris, France}
\affiliation{$^{3}$Institut des Hautes Etudes Scientifique, F-91440 Bures-sur-Yvette, France}
\affiliation{$^{4}$Department of Mathematics, Vanderbilt University, Nashville, TN 37240 USA}

\begin{abstract}
The purpose of this letter is to remove the arbitrariness of the ad hoc choice
of the algebra and its representation in the noncommutative approach to the
Standard Model, which was \emph{begging} for a conceptual explanation. We
assume as before that space-time is the product of a four-dimensional manifold
by a finite noncommmutative space F. The spectral action is the pure
gravitational action for the product space. To remove the above arbitrariness,
we classify the irreducibe geometries F consistent with imposing reality and
chiral conditions on spinors, to avoid the fermion doubling problem, which
amounts to have total dimension $10$ (in the $K$-theoretic sense). It gives,
almost uniquely, the Standard Model with all its details, predicting the
number of fermions per generation to be 16, their representations and the
Higgs breaking mechanism, with very little input. The geometrical model is
valid at the unification scale, and has relations connecting the gauge
couplings to each other and to the Higgs coupling. This gives a prediction of
the Higgs mass of around 170 GeV and a mass relation connecting the sum of the
square of the masses of the fermions to the W mass square, which enables us to
predict the top quark mass compatible with the measured experimental value. We
thus manage to have the advantages of both SO(10) and Kaluza-Klein
unification, without paying the price of plethora of Higgs fields or the
infinite tower of states.

\end{abstract}
\maketitle










\bigskip In the phenomenological approach to determining the Lagrangian of the
fundamental interactions all the present data is consistent with the Standard
Model with neutrino mixing. The input that goes into the construction of the
Standard Model is the following. First one needs the list of three families of
sixteen quarks and leptons and their representations under the gauge group
$SU(3)_{\text{c}}\times SU(2)_{\text{w}}\times U(1)_{\text{Y}}$ . For the
first family, this is taken to be
\[
\left(
\begin{array}
[c]{c}%
u\\
d
\end{array}
\right)  _{L}=\left(  3,2,\frac{1}{3}\right)  _{L},u_{R}=\left(  3,1,\frac
{4}{3}\right)  _{R},d_{R}=\left(  3,1,-\frac{2}{3}\right)  _{R}%
\]
for the quarks and
\[
\left(
\begin{array}
[c]{c}%
\nu\\
e
\end{array}
\right)  _{L}=\left(  1,2,-1\right)  _{L},\,e_{R}=\left(  1,1,-2\right)
_{R},\,\nu_{R}=\left(  1,1,0\right)  _{R}%
\]
for the leptons. The gauge symmetry is then broken to $SU(3)_{\text{c}}\times
U(1)_{\text{em}}$ by employing a complex scalar field Higgs doublet $H$ \ with
representation $\left(  1,2,1\right)  $ $.$ The Lagrangian is constructed by
writing the most general renormalizable interactions consistent with the above
symmetries. The freedom in the choice of the gauge group and the
fermionic\ representations have led to many attempts to unify all the gauge
interactions in one group, and the fermions in one irreducible representation.
The most notable among the unification schemes are models based on the
$SO\left(  10\right)  $ gauge group and groups containing it such as $E_{6},$
$E_{7}$ and $E_{8}.$ The most attractive feature of $SO\left(  10\right)  $ is
that all the fermions in one family fit into the $16$ spinor representation
and the above delicate hypercharge assignments result naturally after the
breakdown of symmetry. However, what is gained in the simplicity of the spinor
representation and the unification of the three gauge coupling constants into
one $SO\left(  10\right)  $ gauge coupling is lost in the complexity of the
Higgs sector. To break the $SO\left(  10\right)  $ symmetry into
$SU(3)_{\text{c}}\times U(1)_{\text{em}}$ one needs to \ employ many Higgs
fields in representations such as $10,$ $120,$ $126$ \cite{Georgi}. The
arbitrariness in the Higgs sector reduces the predictivity of all these models
and introduces many arbitrary parameters, in addition to the unobserved proton decay.

The noncommutative geometric approach \cite{NC} to the unification of all
fundamental interactions, including gravity, is based on the three ansatz
\cite{ACAC}, \cite{ACM}:

\begin{itemize}
\item Space-time is the product of an ordinary Riemannian manifold $M$ by a
finite noncommutative space $F$.

\item The K-theoretic dimension (defined below) of $F$ is $6$ modulo $8.$

\item The physical action functional is given by the spectral action at
unification scale.
\end{itemize}

The empirical data taken as input are:

\begin{itemize}
\item There are $16$ chiral fermions in each of three generations.

\item The photon is massless.

\item There are Majorana mass terms for the neutrinos.
\end{itemize}

Furthermore one makes the following ``ad hoc" choice

\begin{itemize}
\item The algebra of the finite space is taken to be $\mathbb{C}%
\oplus\mathbb{H}\oplus M_{3}\left(  \mathbb{C}\right)  $ where $\mathbb{H}$ is
the algebra of quaternions and $M_{3}\left(  \mathbb{C}\right)  $ is the
algebra of complex $3\times3$ matrices. One of the main purposes of this
letter is to show how this algebra arises.
\end{itemize}

With this input the basic data of noncommutative geometry is constructed,
consisting of an involutive algebra $\mathcal{A}$ of operators in Hilbert
space $\mathcal{H}$, which plays the role of the algebra of coordinates, and a
self-adjoint operator $D$ in $\mathcal{H}$ \cite{NC} which plays the role of
the inverse of the line element. It was shown in \cite{ACM} that the fermions
lie in the desired representations, and that the spectral action associated
with this noncommutative space unifies gravitation with the Standard Model at
the unification scale.

Although the emerging geometrical picture is very appealing, and could be
tested experimentally, the ad hoc choice of the algebra forces us to address
the question of why singling this specific choice, and whether there are other
possibilities, as in the case of grand unification. In addition, taking the
number of fundamental fermions to be $16$ as input, prompts the question of
whether there could exist additional fermions, and whether there is a
mathematical restriction on this number from the representations of the
algebra. It is the purpose of this letter to remove the choice of the algebra
as input, \ and derive it by classifying the possible algebras compatible with
the axioms of noncommutative geometry and minimal number of assumptions to be
specified. We shall keep as physical input that there are three generations,
the photon is massless, and that some of the fermions must acquire a Majorana
mass. We shall prove that the number of fermions must be equal to the square
of an even integer, and thus are able to derive that there are $16$ fermions
per generation. We shall show that the axioms of noncommutative geometry
essentially allows the choice of the algebra to be $\mathbb{C}\oplus
\mathbb{H}\oplus M_{3}\left(  \mathbb{C}\right)  .$ The proof of these results
are rather involved, and we shall only state the theorems, with the rigorous
mathematical details given in \cite{prepare}.

The algebra $\mathcal{A}$ is a tensor product which geometrically corresponds
to a product space. The spectral geometry of $\mathcal{A}$ is given by the
product rule $\mathcal{A}=C^{\infty}\left(  M\right)  \otimes\mathcal{A}_{F}$
\ where the algebra $\mathcal{A}_{F}$ is finite dimensional, and
\[
\mathcal{H}=L^{2}\left(  M,S\right)  \otimes\mathcal{H}_{F},\quad
D=D_{M}\otimes1+\gamma_{5}\otimes D_{F},
\]
where $L^{2}\left(  M,S\right)  $ is the Hilbert space of $L^{2}$ spinors, and
$D_{M}$ is the Dirac operator of the Levi-Civita spin connection on $M.$ The
Hilbert space $\mathcal{H}_{F}$ is taken to include the physical fermions. The
chirality operator is $\gamma=\gamma_{5}\otimes\gamma_{F}.$ The real structure
$J=J_{M}\,\otimes J_{F}$ is an antilinear isometry $J:\mathcal{H}%
\rightarrow\mathcal{H}$ with the property that
\[
J^{\,2}=\varepsilon,\qquad JD=\varepsilon^{\prime}DJ,\qquad J\gamma
=\varepsilon"\gamma J
\]
where $\varepsilon,\varepsilon^{\prime},\varepsilon"\in\left\{  \pm1\right\}
^{3}.$ There are $8$ possible combinations for $\varepsilon,\varepsilon
^{\prime},\varepsilon"$and this defines a K-theoretic dimension of the
noncommutative space mod $8.$ These dimensions are identical to the dimensions
of Euclidean spaces allowing the definitions for Majorana and Weyl spinors. In
order to avoid the fermion doubling problem it was shown in \cite{Alain},
\cite{Barrett}, that the finite dimensional space must be taken to be of
K-theoretic dimension $6$ where in this case $\left(  \varepsilon
,\varepsilon^{\prime},\varepsilon"\right)  =(1,1,-1).$ This makes the total
K-theoretic dimension of the noncommutative space to be $10$ and would allow
to impose the reality (Majorana) condition and the Weyl condition
simultaneously in the Minkowskian continued form, a situation very familiar in
ten-dimensional supersymmetry. In the Euclidean version, the use of the $J$
\ in the fermionic action, would give for the chiral fermions in the path
integral, a Pfaffian instead of determinant, and will thus cut the fermionic
degrees of freedom by 2. In other words, to have the fermionic sector free of
the fermionic doubling problem we \ must make the choice
\[
J_{F}^{\,2}=1,\qquad J_{F}D_{F}=D_{F}J_{F},\qquad J_{F}\,\gamma_{F}%
=-\gamma_{F}J_{F}%
\]
In what follows we will restrict our attention to determination of the finite
algebra, and will omit the subscript $F$ where $F$ stands for Finite.

There are two main constraints on the algebra from the axioms of
noncommutative geometry. We first look for involutive algebras ${\mathcal{A}}$
of operators in ${\mathcal{H}}$ such that,
\[
\lbrack a,b^{0}]=0\,,\quad\forall\,a,b\in{\mathcal{A}}\,.
\]
where for any operator $a$ in ${\mathcal{H}}$, $a^{0}=Ja^{\ast}J^{\,-1}$. This
is called the order zero condition. We shall assume that the following two
conditions to hold. First, the action of ${\mathcal{A}}$ has a separating
vector. Second, the representation of ${\mathcal{A}}$ and $J$ \ in
${\mathcal{H}}$ is irreducible.

The strategy to determine the finite space $F$ \ then involves the following
steps. \ First, to classify the irreducible triplets $\left(  \mathcal{A}%
,\mathcal{H},J\right)  .$ Second to impose the $\mathbb{Z}/2$ grading on
$\mathcal{H}.$ Third, to classify all the subalgebras $\mathcal{A}_{F}%
\subset\mathcal{A}$ which allows for an operator $D$ that does not commute
with the center of $\mathcal{A}$ but fulfills the order one condition%
\[
\lbrack\lbrack D,a],b^{0}]=0\qquad\forall\,a,b\in\mathcal{A}_{F}\,.
\]

\bigskip Starting with the classification of the order zero condition with the
irreducible pair $\left(  \mathcal{A},J\right)  $ one finds out that the
solutions fall into two classes. Let $\mathcal{A}_{\mathbb{C}}$ be the complex
linear space generated by by $\mathcal{A}$ in ${\mathcal{L}}({\mathcal{H}})$,
the algebra of operators in ${\mathcal{H}}.$ Then the two classes correspond to

\begin{itemize}
\item The center $Z\left(  \mathcal{A}_{\mathbb{C}}\right)  $ is $\mathbb{C}.$

\item The center $Z\left(  \mathcal{A}_{\mathbb{C}}\right)  $ is
$\mathbb{C\oplus C}.$
\end{itemize}

\section{\bigskip The case $Z\left(  \mathcal{A}_{\mathbb{C}}\right)  =$
$\mathbb{C}$}

In this case we can state the following theorem.

\begin{thm}
Let $\mathcal{H}$ be a Hilbert space of dimension $n$. Then an irreducible
solution with $Z\left(  \mathcal{A}_{\mathbb{C}}\right)  =$ $\mathbb{C}$
exists iff $n=k^{2}$ is a square. It is given by $\mathcal{A}_{\mathbb{C}%
}=M_{k}\left(  \mathbb{C}\right)  $ acting by left multiplication on itself
and antilinear involution
\[
J\left(  x\right)  =x^{\ast},\quad\forall x\in M_{k}\left(  \mathbb{C}\right)
.
\]

\end{thm}

This determines $\mathcal{A}_{\mathbb{C}}$ and its representations in $\left(
\mathcal{A},J\right)  $ and allows only for three possibilities for
$\mathcal{A}$. These are $\mathcal{A=}M_{k}\left(  \mathbb{C}\right)  ,$
$M_{k}\left(  \mathbb{R}\right)  $ and $M_{a}\left(  \mathbb{H}\right)  $ for
even $k=2a,$ where $\mathbb{H}$ is the field of quaternions. These correspond
respectively to the unitary, orthogonal and symplectic case.

\medskip

\textbf{{$\mathbb{Z}/2-$Grading}}

\medskip

In the set up of spectral triples one assumes that in the even case the
Hilbert space $\mathcal{H}$ is $\mathbb{Z}/2$ -graded, i.e. endowed with a
grading operator $\gamma=\gamma^{\ast},$ $\gamma^{2}=1$ such that
$\gamma\mathcal{A}\gamma^{-1}=\mathcal{A}$. In the $Z\left(  \mathcal{A}%
_{\mathbb{C}}\right)  =$ $\mathbb{C}$ case, one can then show that it is not
possible to have the finite space to be of \ K-theoretic dimension $6$, with
$J\gamma=-\gamma J.$ We therefore can proceed directly to the second case.

\section{The case $Z\left(  \mathcal{A}_{\mathbb{C}}\right)  =$
$\mathbb{C\oplus C}$}

\bigskip In this case we can state the theorem

\begin{thm}
Let \ $\mathcal{H}$ be a Hilbert space of dimension $n$. Then an irreducible
solution with $Z\left(  \mathcal{A}_{\mathbb{C}}\right)  =$ $\mathbb{C\oplus
C}$ exists iff $n=2k^{2}$ is twice a square. It is given by $\mathcal{A}%
_{\mathbb{C}}=M_{k}\left(  \mathbb{C}\right)  \oplus M_{k}\left(
\mathbb{C}\right)  $ acting by left multiplication on itself and antilinear
involution
\[
J\left(  x,y\right)  =\left(  y^{\ast},x^{\ast}\right)  ,\quad\forall x,y\in
M_{k}\left(  \mathbb{C}\right)  .
\]

\end{thm}

With each of the $M_{k}\left(  \mathbb{C}\right)  $ in $\mathcal{A}%
_{\mathbb{C}}$ we can have the three possibilities $M_{k}\left(
\mathbb{C}\right)  ,$ $M_{k}\left(  \mathbb{R}\right)  ,$ or $M_{a}\left(
\mathbb{H}\right)  ,$ where $k=2a$. At this point we make the
\textit{hypothesis} that we are in the ``symplectic--unitary" case, thus
restricting the algebra $\mathcal{A}$ to the form $\mathcal{A}=M_{a}\left(
\mathbb{H}\right)  \oplus M_{k}\left(  \mathbb{C}\right)  ,$ $k=2a.$ The
dimension of the Hilbert space $n=2k^{2}$ then corresponds to $k^{2}$
fundamental fermions, where $k=2a$ is an even number. The first possible
\ value for $k$ is $2$ corresponding to a Hilbert space of four fermions and
an algebra $\mathcal{A}=\mathbb{H}\oplus M_{2}\left(  \mathbb{C}\right)  $.
The existence of quarks rules out this possibility. The next possible value
for $k$ is $4$ predicting the number of fermions to be $16.$

\medskip

\textbf{{$\mathbb{Z}/2-$Grading}}

\medskip

In the above symplectic--unitary case, one can write the Hilbert space
$\mathcal{H}$ as the sum of the spaces of $\mathbb{C}$-linear maps from $V$ to
$W$ and from $W$ to $V$ where $V$ is a $4$-dimensional vector space over
$\mathbb{C}$ and $W$ a $2$-dimensional right vector space over $\mathbb{H}$.
There exists, up to equivalence, a unique $\mathbb{Z}/2-$grading of $W$ and it
induces uniquely a $\mathbb{Z}/2-$grading $\gamma$ of $\mathcal{E}$. One then
takes the grading $\gamma$ of $\mathcal{H}$ so that the K-theoretic dimension
of the finite space is $6,$ which means that $J\,\gamma=-\gamma J.$ It is
given by
\[
\gamma\left(  \zeta,\eta\right)  =\left(  \gamma\zeta,-\gamma\eta\right)
\]
This grading breaks the algebra $\mathcal{A=}$ $M_{2}\left(  \mathbb{H}%
\right)  \oplus M_{4}\left(  \mathbb{C}\right)  $, which is non trivially
graded only for the $M_{2}\left(  \mathbb{H}\right)  $ component, to its even
part:
\[
\mathcal{A}^{\text{ev}}=\text{ }\mathbb{H}\oplus\mathbb{H}\oplus M_{4}\left(
\mathbb{C}\right)  \,.
\]

\medskip

\section{The subalgebra and the order one condition}

From the previous analysis, it should be clear that the only relevant case to
be subjected to the order one condition is $Z\left(  \mathcal{A}_{\mathbb{C}%
}\right)  =$ $\mathbb{C\oplus C}$ and for $\mathcal{A=}$ $M_{2}\left(
\mathbb{H}\right)  \oplus M_{4}\left(  \mathbb{C}\right)  $. The center of the
algebra $Z\left(  \mathcal{A}\right)  $ is non-trivial, and thus the
corresponding space is not connected. The Dirac operator must connect the two
pieces non-trivially, and therefore must satisfy
\[
\left[  D,Z\left(  \mathcal{A}\right)  \right]  \neq\left\{  0\right\}
\]
The physical meaning of this constraint, is to allow some of the fermions to
acquire Majorana masses, realizing the sea-saw mechanism, and thus connecting
the fermions to their conjugates. The main constraint on such Dirac operators
arises from the order one condition. We have to look for subalgebras
$\mathcal{A}_{F}\subset\mathcal{A}^{\text{ev}},$ the even part of the algebra
$\mathcal{A}$ for which $[[D,a],b^{0}]=0,\quad\forall\,a,b\in\mathcal{A}_{F}$.
We can now state the main result which recovers the input of \cite{ACM}.

\begin{thm}
Up to an automorphisms of $\mathcal{A}^{\text{ev}},$ there exists a unique
involutive subalgebra $\mathcal{A}_{F}\subset\mathcal{A}^{\text{ev}}$ of
maximal dimension admitting off-diagonal Dirac operators. It is given by
\begin{align*}
\mathcal{A}_{F}  &  =\left\{  \lambda\oplus q,\lambda\oplus m\,|\lambda
\in\mathbb{C},q\in\mathbb{H},m\in M_{3}\left(  \mathbb{C}\right)  \right\} \\
&  \subset\mathbb{H}\oplus\mathbb{H}\oplus M_{4}\left(  \mathbb{C}\right)
\end{align*}
using a field morphism $\mathbb{C}\rightarrow\mathbb{H},$ The involutive
algebra $\mathcal{A}_{F}$ is isomorphic to $\mathbb{C}\oplus\mathbb{H}\oplus
M_{3}\left(  \mathbb{C}\right)  $ and together with its representation in
$\left(  \mathcal{H},J,\gamma\right)  $ gives the noncommutative space taken
as input in \cite{ACM}.
\end{thm}

In simple terms, this means that the off-diagonal elements of the Dirac
operator, connecting the $16$ spinors to their conjugates, break
$\mathbb{H}\oplus\mathbb{H}\oplus M_{4}\left(  \mathbb{C}\right)  \rightarrow$
$\mathbb{C}\oplus\mathbb{H}\oplus M_{3}\left(  \mathbb{C}\right)  .$ We have
thus recovered the main input used in deriving the standard model with the
minimal empirical data. These are the masslessness of the photon and the
existence of mixing terms for fermions and their conjugates. The main
mathematical inputs are that the representations of $\left(  \mathcal{A}%
,J\right)  $ are irreducible, there is an anti-linear isometry with
non-trivial grading on one of the algebras. Having made contact with the
starting point of \cite{ACM} we summarize the results in that work.

Let $M$ be a Riemannian spin $4$ -manifold and $F$ \ the finite noncommutative
geometry of K-theoretic dimension $6$ but with multiplicity $3.$ Let
\ $M\times F$ \ be endowed with the product metric. The unimodular subgroup of
the unitary group acting by the adjoint representation Ad$\left(  u\right)  $
in $\mathcal{H}$\ is the group of gauge transformations $SU(2)_{\text{w}%
}\times U(1)_{\text{Y}}\times SU(3)_{\text{c}}.$ The unimodular inner
fluctuations of the metric give the gauge bosons of SM. The full standard
model (with neutrino mixing and seesaw mechanism) coupled to gravity is given
in Euclidean form by the action functional \cite{ACAC}, \cite{ACM}%
\[
S=\text{Tr}\left(  f\left(  \frac{D_{A}}{\Lambda}\right)  \right)  +\frac
{1}{2}\left\langle \,J\,\widetilde{\xi},D_{A}\widetilde{\xi}\right\rangle
,\quad\widetilde{\xi}\in\mathcal{H}_{\text{cl}}\quad
\]
where $D_{A}$ is the Dirac operator with inner fluctuations. To explain the
role of the spectral action principle, we note that one of the virtues of the
axioms of noncommutative geometry is that it allows for a shift of point of
view, similar to Fourier transform in which the usual emphasis on the points
$x\in M$ \ of a geometric space is replaced with the spectrum $\Sigma
\subset\mathbb{R}$ of the operator $D.$ The hypothesis which is stronger than
diffeomorphism invariance is that "\emph{The physical action only depends upon
}$\Sigma$".

We conclude that our approach predicts a unique fermionic representation of
dimension $16,$ with gauge couplings unification. These properties are only
shared with the $SO(10)$ grand unified theory. The main advantage of our
approach over the grand unification approach is that the reduction to the
Standard Model gauge group is not due to plethora of Higgs fields, but is
naturally obtained from the order one condition, which is one of the axioms of
noncommutative geometry. There is also no proton decay because there are no
additional vector particles linking the lepton and quark sectors. The spectral
action is the pure gravitational sector of the noncommutative space. This is
similar in spirit to the Kaluza-Klein approach, but with the advantage of
having a finite spectrum, and not the infinite tower of states. Thus the
noncommutative geometric approach manages to combine the advantages of both
grand unification and Kaluza-Klein without paying the price of introducing
many unwanted states. We still have few delicate points which require further
understanding. The first is to understand the need for the restriction to the
symplectic--unitary case which is playing an important role in the
construction. The second is to determine the number of generations$.$ From the
physics point, because of CP\ violation, we know that we need to take $N\geq
3$, but there is no corresponding convincing mathematical principle.

We would like to stress that the spectral action of the standard model comes
out almost uniquely, predicting the number of fermions, their representations
and the Higgs breaking mechanism, with very little input. The geometrical
model is valid at the unification scale, and relates the gauge coupling
constants to each other and to the Higgs coupling. When these relations are
taken as boundary conditions valid at the unification scale in the
renormalization group (RG) equations, one gets a prediction of the Higgs mass
to be around $170\pm10$ GeV, the error being due to our ignorance of the
physics at unification scale. In addition there is one relation between the
sum of the square of fermion masses and the $W$ particle mass square%
\[%
{\displaystyle\sum\limits_{\text{generations}}}
\left(  m_{e}^{2}+m_{\nu}^{2}+3m_{u}^{2}+3m_{d}^{2}\right)  =8M_{W}^{2}.
\]
which enables us to predict the top quark mass compatible with the measured
experimental value.

We note that general studies of the Higgs sector in the standard model
\ \cite{Herbi} show that when the Higgs and top quark masses are comparable,
as in our case, \ then the Higgs mass will be stable under the renormalization
group equations, up to the Planck scale.

\medskip

\begin{acknowledgments}
The research of A. H. C. is supported in part by the National Science
Foundation under Grant No. Phys-0601213 and by a fellowship from the Arab Fund
for Economic and Social Development. He would like to thank \ Herbi Dreiner
for pointing reference \cite{Herbi} to him.
\end{acknowledgments}

\end{document}